\documentclass[]{emulateapj}
\usepackage{graphicx,natbib}
\begin{document}

\newcommand{\ftd}{F_{\rm td}}
\newcommand{\tc}{T_{\rm c}}
\newcommand{\teff}{T_{\rm eff}}
\newcommand{\fc}{f_{\rm c}}
\newcommand{\ratio}{(R_{\rm app}/D)^2}
\newcommand{\fortyeight}{SAX~J1748.9$-$2021}
\newcommand{\twenty}{4U~1820$-$30}
\newcommand{\fortyfive}{EXO~1745$-$248}
\newcommand{\thirtyone}{KS~1731$-$260}
\newcommand{\twentyfour}{4U~1724$-$207}
\newcommand{\oeight}{4U~1608$-$52}
\newcommand{\actaa}{Acta Astronomica}

\title{Data Selection Criteria for Spectroscopic Measurements of Neutron
Star Radii with X-ray Bursts}

\author{Feryal \"Ozel\altaffilmark{1}, Dimitrios
  Psaltis\altaffilmark{1}, Tolga G\"uver\altaffilmark{2} }

\email{E-mail: fozel@email.arizona.edu}

\altaffiltext{1}{Department of Astronomy, University of Arizona, 933
  N. Cherry Ave., Tucson, AZ 85721, USA}

\altaffiltext{2}{Istanbul University, Science Faculty, Department of
  Astronomy and Space Sciences, Beyazit, 34119, Istanbul, Turkey}

\begin{abstract}
Data selection and the determination of systematic uncertainties in
the spectroscopic measurements of neutron star radii from
thermonuclear X-ray bursts have been the subject of numerous recent
studies. In one approach, the uncertainties and outliers were
determined by a data-driven Bayesian mixture model, whereas in a
second approach, data selection was performed by requiring that the
observations follow theoretical expectations. We show here that, due
to inherent limitations in the data, the theoretically expected trends
are not discernible in the majority of X-ray bursts even if they are
present. Therefore, the proposed theoretical selection criteria are
not practical with the current data for distinguishing clean data sets
from outliers. Furthermore, when the data limitations are not taken
into account, the theoretically motivated approach selects a small
subset of bursts with properties that are in fact inconsistent with
the underlying assumptions of the method. We conclude that the
data-driven selection methods do not suffer from the limitations of
this theoretically motivated one.
\end{abstract}


\section{Introduction}

Thermonuclear flashes on neutron stars have been used in the past
decade to perform spectroscopic measurements of neutron star radii and
masses (e.g., \citealt{majczyna2005,ozel2006,ozel2009a,ozel2010a,poutanen2014}).
The rich burst dataset \citep{galloway2008} from the Rossi X-ray
Timing Explorer (RXTE) has not only allowed the measurement of the
macroscopic properties of half a dozen of neutron stars but also
initiated detailed studies of systematic uncertainties in these
measurements. Even though the majority of sources showed bursting
behavior that is highly reproducible \citep{guver2012a,guver2012b}, a
subset shows burst properties and evolution that are more complex.

A number of studies explored different approaches to identifying
outliers in the burst samples of different sources that contaminate
the statistical inferences in the measurements. In
\citet{guver2012a,guver2012b}, we used a data-driven approach that employs
a Gaussian mixture Bayesian inference to reject outliers. In
particular, we looked at the cooling tails of X-ray bursts in the
flux-temperature diagram, which would have yielded identical tracks
among bursts of the same source in the absence of any astrophysical
complexities. The degree of scatter we observed in the cooling tracks
allowed us to measure the level of systematic uncertainty in the
measurements, e.g., due to obscuration, reflection off the accretion
disk, or uneven burning on the stellar surface, as well as to remove a
small percentage of cooling tracks that clearly did not behave like
the majority. 

In an alternate approach, Poutanen et al.\ (2014; see also
\citealt{kajava2014}) developed a theoretically motivated method to
select bursts. They used the bursting neutron star atmosphere models
of \citet{suleimanov2012} to calculate the evolution of the blackbody
normalization and chose only those bursts that followed their
predicted trends. When applied to the source \oeight, they found that
this approach selects a very small fraction of bursts as acceptable
ones and leads to substantially larger inferred neutron star radii.

In this paper, we show that the bursts selected by the data-driven
approach of \citet{guver2012a,guver2012b} are not in conflict with the
theoretical expectations of the \citet{suleimanov2012} atmosphere
models. The theoretical models predict that the blackbody
normalization evolves rapidly in a narrow range of fluxes close to the
Eddington limit. However, in most bursts, the instrument limitations
of RXTE do not allow resolving this fast predicted evolution. When
this limitation is not taken into account in burst selection, as was
the case in \citet{poutanen2014}, this procedure biases the selection
toward a peculiar set of bursts. We further show that this subset of
bursts are not photospheric radius expansion bursts, which is a
requirement in the theoretical motivated selection criteria of
\citet{poutanen2014} and \citet{kajava2014}. Because of both its
practical limitations and its biased results, we argue that this data
selection method does not lead to reliable radius measurements.

\begin{figure}
\centering
\includegraphics[scale=0.38]{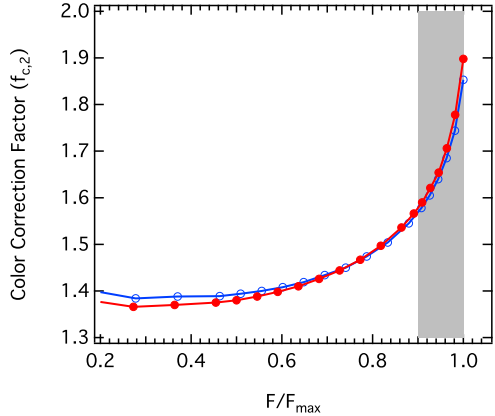}
\caption{The dependence of the color-correction factor in two He
  models (\# 17 and \#18) of \citet{suleimanov2012} as a function
  of the burst luminosity divided by its maximum value. The vertical
  grey band corresponds to a 10\% flux change, which cannot be
  resolved for the vast majority of radius expansion bursts observed
  with RXTE (see Figure~\ref{fig:flux_distrib}). Even though the
  models predict color correction factors as large as 1.9, the time
  resolution used in the analysis restricts the maximum effective
  color correction factor to $\lesssim 1.6$.}
\label{fig:color_corr}
\end{figure}

\section{Applying the Theoretical Selection Criteria to RXTE Bursts}

In theoretical models of bursting neutron star atmospheres, the deviation 
of the spectrum from a blackbody is typically quantified in terms of 
the color correction factor 
\begin{equation}
f_{\rm c} \equiv T_{\rm c} / T_{\rm eff}, 
\end{equation}
where $\tc$\ is the color temperature obtained from the spectral fits
and $\teff$\ is the effective temperature of the atmosphere. The color
correction factor depends on the effective temperature of the
atmosphere, its composition, and the effective gravitational
acceleration, i.e., how close the radiation flux is to the Eddington
critical value. Because $\fc \ne 1$ in all relevant conditions, the 
normalization of the blackbody, defined as 
\begin{equation} 
K \equiv (R_{\rm BB}/D)^2 = \frac{F}{\sigma_B \tc^4}, 
\label{eq:Kdef}
\end{equation}
where $F$ is the flux of the thermal emission and $\sigma_B$ is the 
Stefan-Boltzmann constant, is not equal to the apparent angular 
size of the neutron star, $\ratio$, but rather to 
\begin{equation}
K = \fc^{-4} \ratio.
\label{eq:Kfc}
\end{equation}
In Figure~\ref{fig:color_corr}, we plot a typical evolution of the
color correction factor for two of the models described in
\citet{suleimanov2012}. It is evident in this figure that most of the
evolution of the color correction factor occurs within $10\%$ of the
maximum flux (i.e., the Eddington flux corrected for temperature
effects).

\begin{figure}
\centering
\includegraphics[scale=0.38]{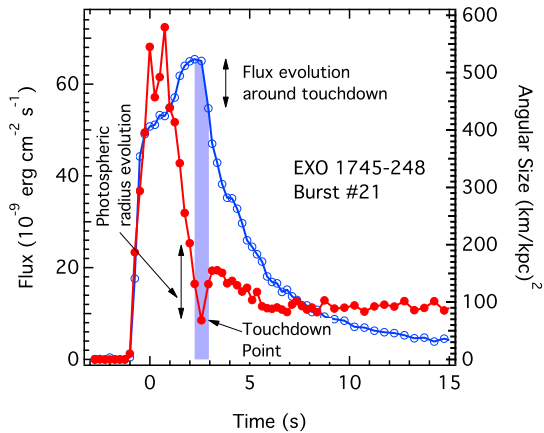}
\caption{The evolution of the X-ray flux {\em (blue)\/} and of the
  apparent angular size {\em (red)\/} during burst \#21 from
  EXO~1745$-$52. In this case, the X-ray flux evolves by 18\% between
  the two time bins adjacent to the touchdown point.}
\label{fig:1745_ex}
\end{figure}

For a given neutron star, $\ratio$\ is fixed, and therefore, the
blackbody normalization is expected to scale as $\fc^4$. Ideally, this
expected evolution of the blackbody normalization can be used as a
selection criterium \citep{poutanen2014,kajava2014} as well as a way
of measuring a combination of the stellar mass and radius
\citep{majczyna2005a,poutanen2014}.

Unfortunately, a number of complexities make the application of this
selection procedure to RXTE data impractical. First, this method
requires the selecting bursts that reach or exceed the Eddington
limit, which is usually achieved by searching for evidence for
photosperic radius expansion (PRE) episodes. Second, it requires
accurately pinpointing the time of touchdown at the end of the PRE
event and using only the data after this point to measure the
evolution of the blackbody normalization. Finally, it requires
obtaining a large number of counts at short integration times in order
to resolve the evolution of the blackbody normalization with the
rapidly decreasing flux. 

\begin{figure}
\centering
\includegraphics[scale=0.35]{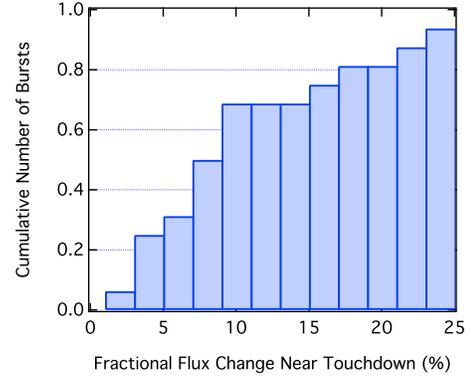}
\includegraphics[scale=0.35]{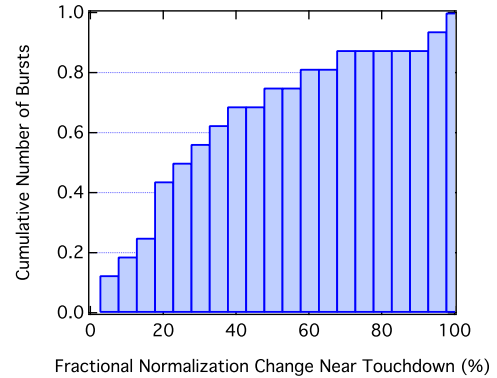}
\caption{{\em (Top)\/} The cumulative distribution of the fractional
  burst flux evolution between the two 0.25~s time bins that are
  adjacent to each touchdown point, for all the photospheric radius
  expansion bursts used in \citet{ozel2015} for radius
  measurements. {\rm (Bottom)\/} The cumulative distribution of the
  fractional change in the blackbody normalization during the 0.25~s
  time bin prior to each touchdown point, for all the photospheric
  radius expansion bursts used in the \citet{ozel2015} study.  In the
  majority of bursts, the flux evolves by 5-20\% and the blackbody
  normalization by 20-80\% around the touchdown point, in a way that
  cannot be resolved with current observations.}
\label{fig:flux_distrib}
\end{figure}

We have discussed the criteria to select bona fide PRE bursts in
\citet{guver2012a} and will revisit this point in the next section.
Because of the burst countrates and instrumental limitations, burst
spectra with RXTE are extracted using $\sim 0.25$~s integrations.
Using a typical PRE burst from \fortyfive, we show in
Figure~\ref{fig:1745_ex} how this time binning affects both the time
localization of the touchdown point and our ability to resolve the
rapid flux evolution after it. In this example, within one 0.25~s time
bin, the flux to the right of the touchdown point evolves by 17\%,
while the photospheric radius to the left of the touchdown point
evolves by 63\%. As can be seen from eq.~(\ref{eq:Kfc}) and taking
into account the dependence of $\fc$\ on the flux (see
Fig.\ref{fig:color_corr}), these two changes have opposite effects on
the blackbody normalization, effectively canceling each other out.

\begin{figure}
\centering \includegraphics[scale=0.38]{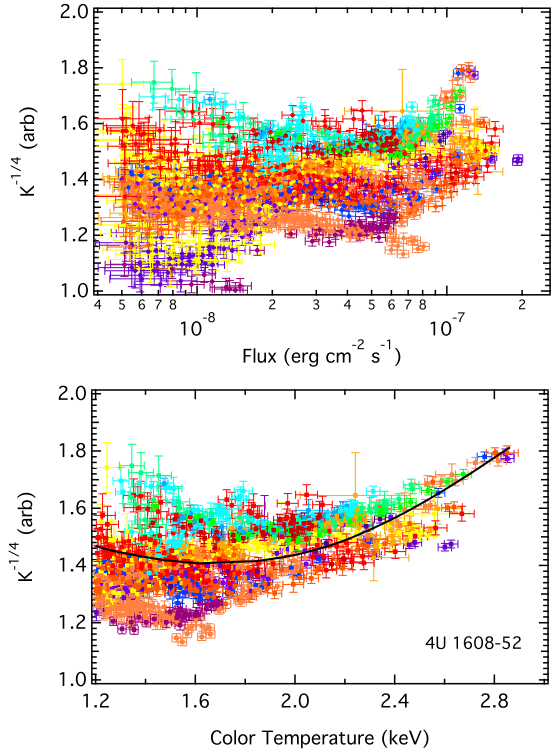}
\caption{The inferred evolution of the color correction factor for all
  the bursts from 4U~1608$-52$ (modulo an arbitrary normalization), as
  a function of {\em (top)} the measured X-ray flux and {\em (bottom)}
  the color temperature of the spectrum. Different colors correspond
  to different bursts. In the bottom panel, the evolution of the color
  correction factor from a He-atmosphere model of Suleimanov et
  al.\ (2012) is overplotted for comparison. Plotting the blackbody
  normalization as a function of the color temperature reduces the
  scatter and allows the theoretically expected correlation to start
  becoming apparent.}
\label{fig:color_corr_1608}
\end{figure}

This behavior is not unique to the example shown in
Figure~\ref{fig:1745_ex}. Indeed, an analysis of all the PRE bursts
for the six sources used in the radius measurements of
\citet{ozel2015} shows a similar level of evolution. We show this in
Figure~\ref{fig:flux_distrib}, where we plot in the top panel the
cumulative distribution of the fractional change in the flux at the
two 0.25~s time bins that are adjacent to the touchdown point.  In the
bottom panel, we also show the fractional change in the blackbody
normalization during the 0.25~s time bin prior to the touchdown point.
The typical values of the former are in the $5-20\%$ range, while for
the latter, in the $20-80\%$ range. As a result, in one $0.25$~s time
bin, the flux typically changes by $\sim 10\%$ after the true
touchdown point, leading to a drop in the color correction factor from
$\sim 1.9$ to $1.6$, and a corresponding change in the blackbody
normalization by $\approx 50\%$. During the same time bin but prior to
the touchdown point, the $\ratio$\ changes also by $\approx 50\%$, in
the opposite direction. Therefore, it becomes impossible to detect the
expected rapid evolution of the blackbody normalization as predicted
by theoretical models.

Looking for the evolution of the blackbody normalization as a function
of flux fails not only on a burst-by-burst basis but also when it is
applied to all the bursts from a given source. Fundamentally, the
color correction factor for a given source depends on the local flux
in the atmosphere, as measured by the effective temperature, and not
on the total observed flux. The latter is affected by a number of
factors, including the time bin averaging discussed above, as well as
effects related to, e.g., partial obscuration of the neutron star
surface or any reflection off of the accretion disk which are thought
to be responsible for the outliers observed in some sources.  This is
of particular importance for the few sources that show a substantial
number of outliers. The same effects also introduce a scatter to the
blackbody normalization. Therefore, when the quantity $K^{-1/4}$ is
plotted as a function of the measured flux, as shown in
Figure~\ref{fig:color_corr_1608} for the case of \oeight, the
theoretically expected dependence is masked by the scatter. If we
plot, instead, the quantity $K^{-1/4}$ against the observed color
temperature, the scatter in the abcissa is reduced to a point where
the theoretically expected correlation starts becoming evident.

\begin{figure*}
\centering
\includegraphics[scale=0.35]{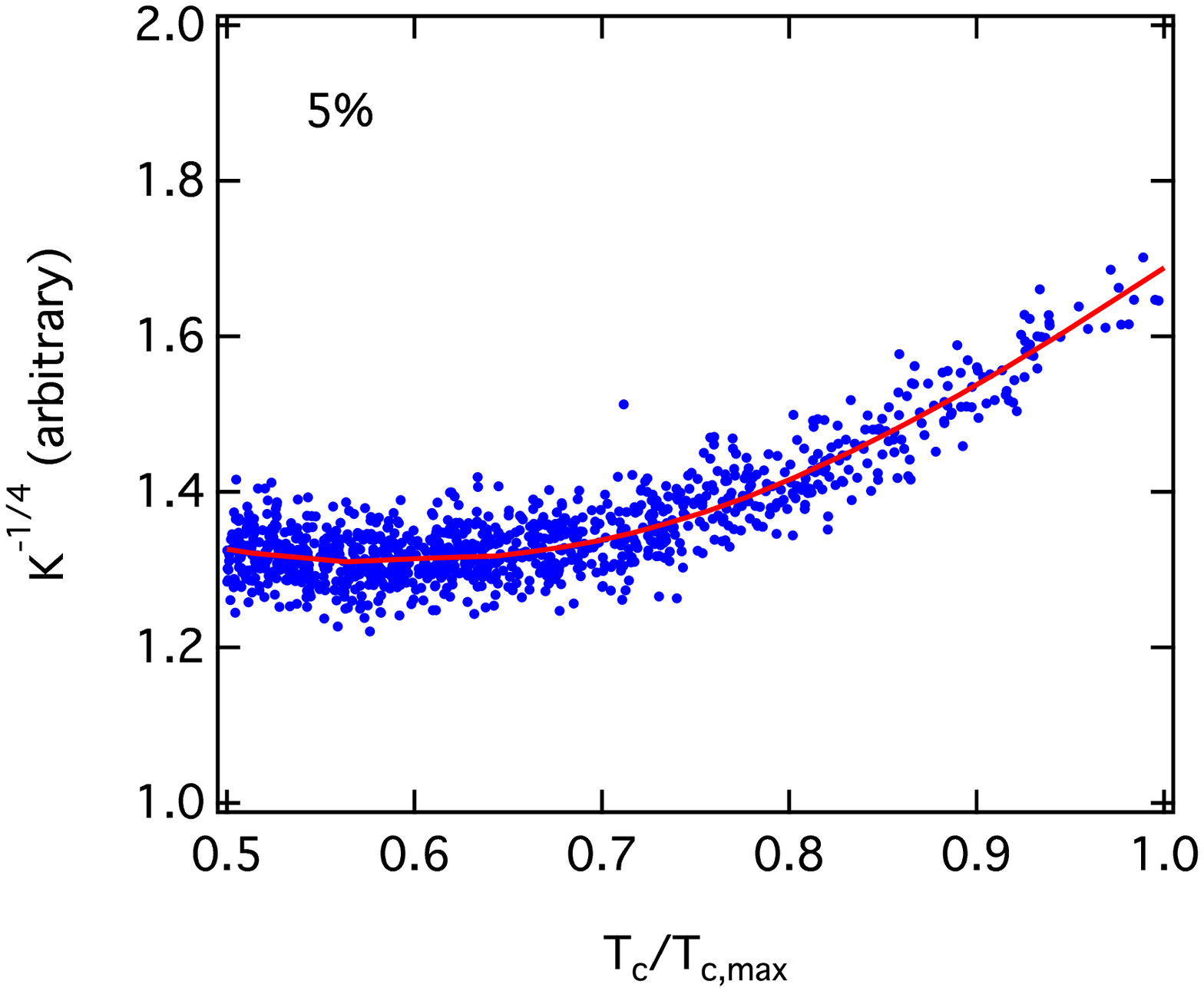}
\includegraphics[scale=0.35]{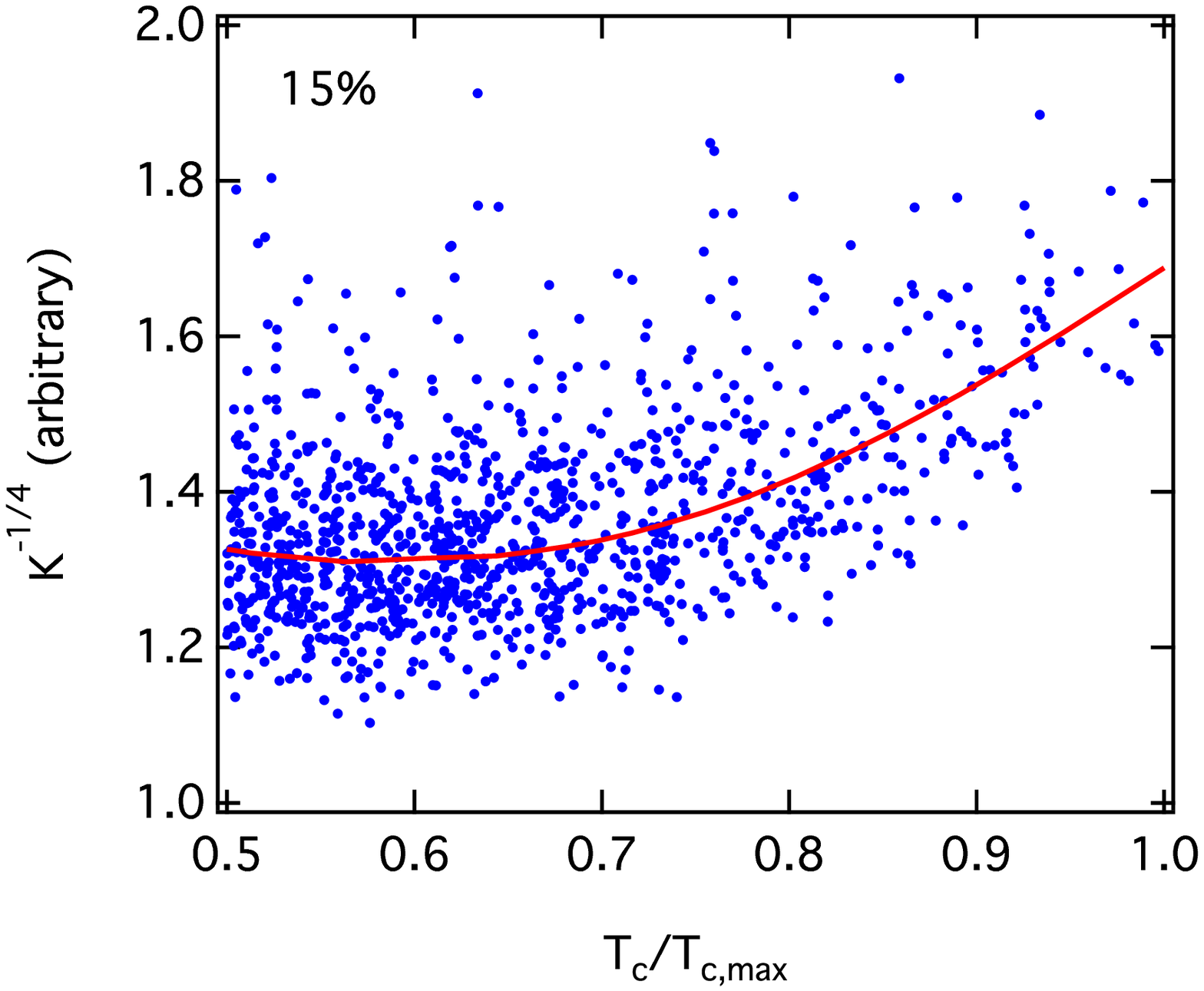}
\includegraphics[scale=0.35]{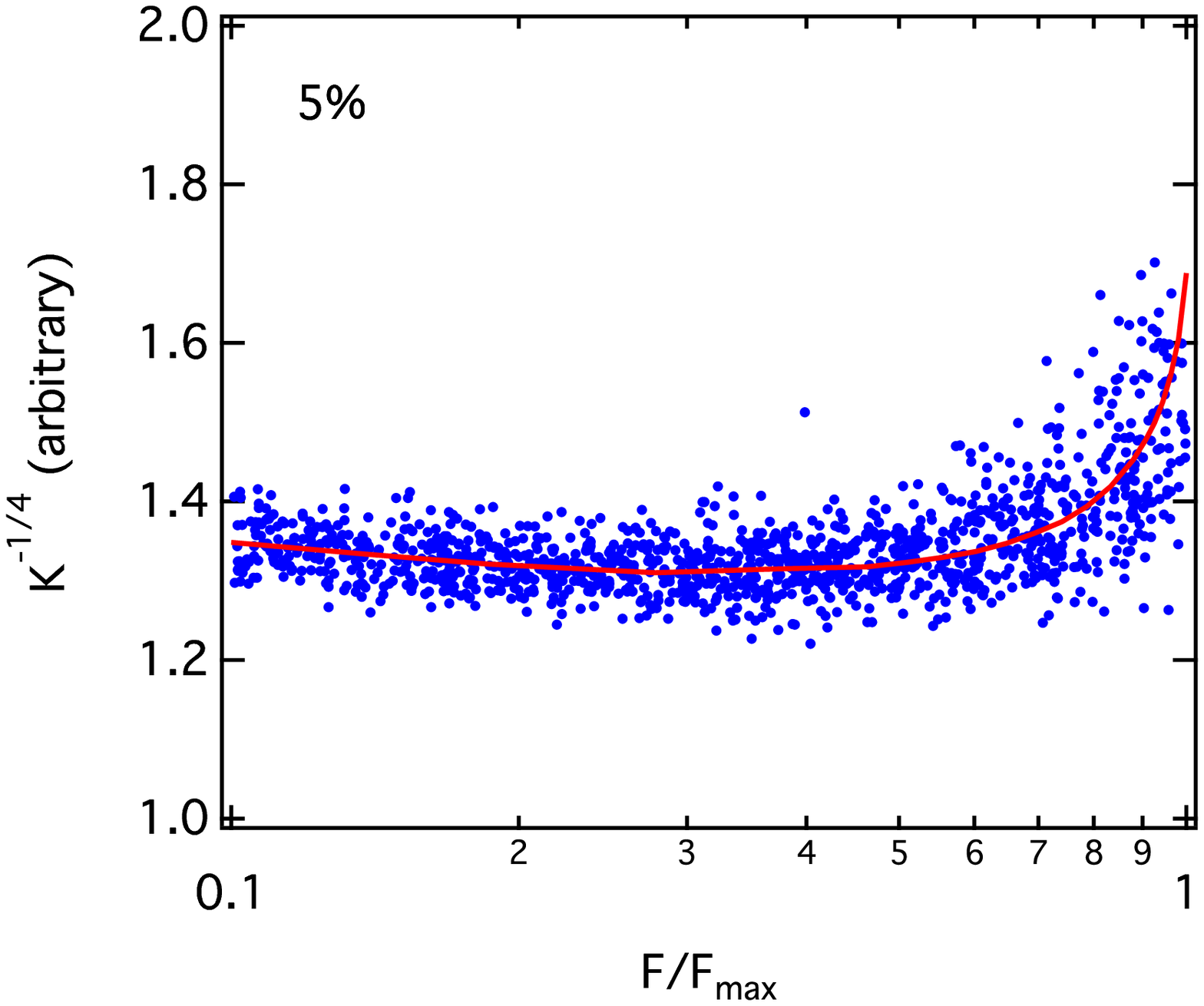}
\includegraphics[scale=0.35]{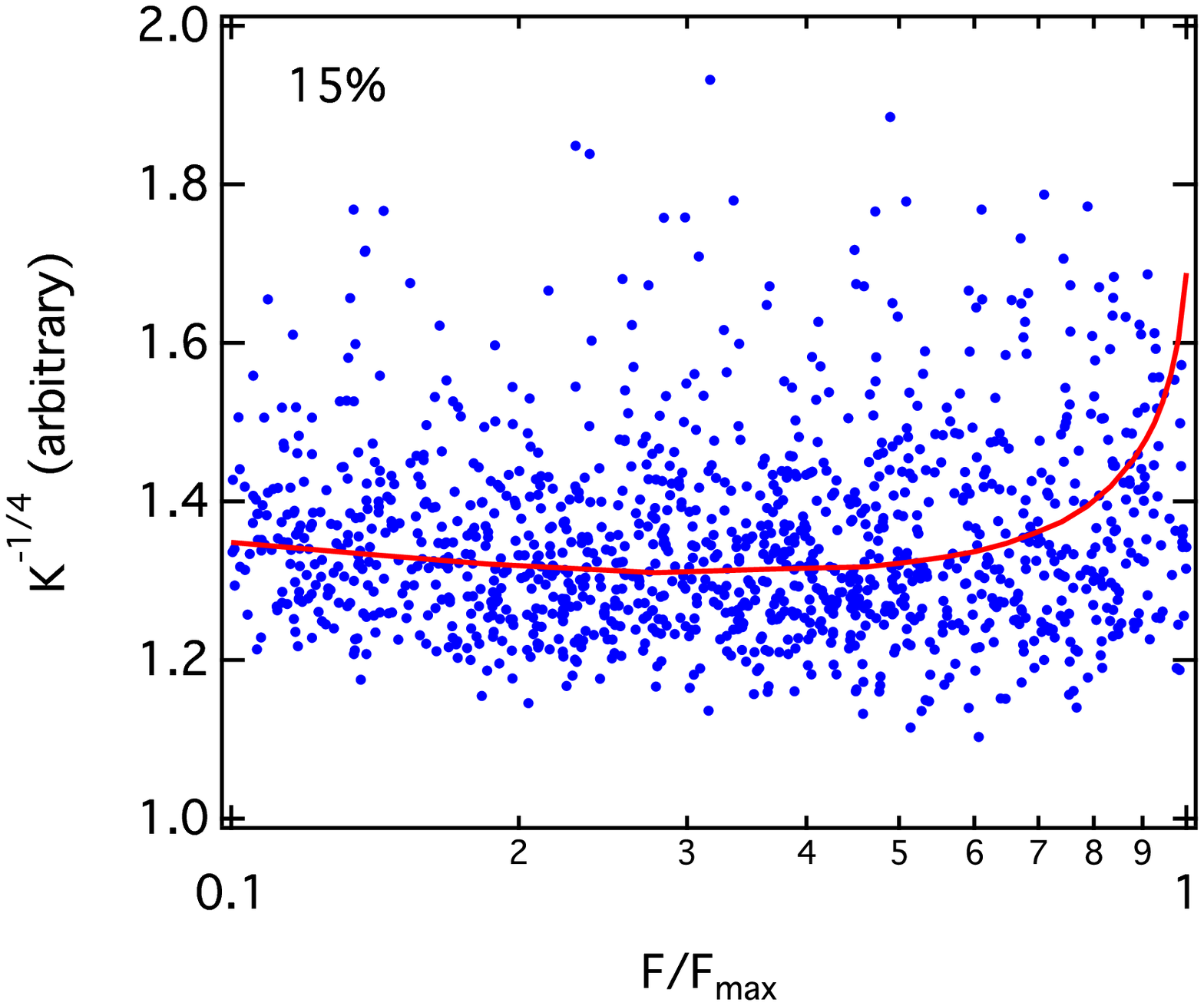}
\caption{Monte Carlo simulations of the correlations between the
  blackbody normalization, color temperature, and flux based on a
  theoretical model of a bursting neutron star atmosphere. The
  theoretical model dependences, from which the mock observations were
  drawn, are shown by the red curves. In the left panels, we assigned
  a 5\% Gaussian spread in the radius of the emitting region, whereas
  in the right panels, we assigned a 15\% spread. Even for the small
  spread, the blackbody normalization appears a lot more tightly
  correlated with color temperature than with flux. For the larger
  spread, the latter correlation disappears and would have given the
  false impression that the observations are inconsistent with
  theoretical expectations. }
\label{fig:monte_carlo}
\end{figure*}

In order to demonstrate the degree of scatter that even a small spread
in the emitting radius can introduce to the expected correlations, we
performed the following Monte Carlo simulation.  We simulated mock
observations of burst spectra using Model \#17 of
\citet{suleimanov2012} for the dependence of the color correction
factor on the effective temperature, which corresponds to a helium
atmosphere with surface gravity $\log g = 14.3$. We assigned a spread
in the radius of the emitting region with a Gaussian likelihood.  We
used these mock observations to infer the values of the blackbody
normalization, color temperature, and flux that would have been
measured under these conditions. Figure~\ref{fig:monte_carlo} shows
the resulting relations between blackbody normalization, color
temperature, and flux, for two cases: the left columns with a 5\%
spread and the right columns with a 15\% spread in the emitting
radius. Even for the small spread, the blackbody normalization appears
a lot more tightly correlated with color temperature than with flux.
For the larger spread, the latter correlation disappears and would
have given the impression that the observations are inconsistent with
theoretical expectations. Note that in performing these simulations,
we have not included the significant smearing that occurs due to the
size of the time bins that we discussed above or the statistical
uncertainties that lead to correlated values between the blackbody
normalization and the blackbody temperature (see, e.g., Figs. $3-8$ of
\citealt{ozel2015}).

The above discussion demonstrates that the bursts selected in the
data-driven approach of \citet{guver2012a,guver2012b} show a spectral
evolution that do not contradict theoretical expectations and can,
therefore, lead to an unbiased sample for the measurement of neutron
star radii (as reported by \citealt{ozel2015}). Moreover, as we will
show in the next section, devising selection criteria based on this
theoretical expectation, without taking into account the inherent
limitations of the data, leads to selecting precisely those bursts
that do not reach the Eddington limit, therefore, violating the basic
assumption of the procedure.

\section{Problems with the Radius Measurements in the Theoretically Selected Bursts}

\citet{poutanen2014} applied the criteria based on the theoretical
evolution of the blackbody normalization to select bursts for the
measurement of the neutron star radius in \oeight. Specifically, they
chose bursts where the blackbody normalization at half the Eddington
flux is at least twice as large as the blackbody normalization at what
they marked as the touchdown point in their sample of Eddington
limited bursts. 

\begin{figure*}
\centering \includegraphics[scale=0.35]{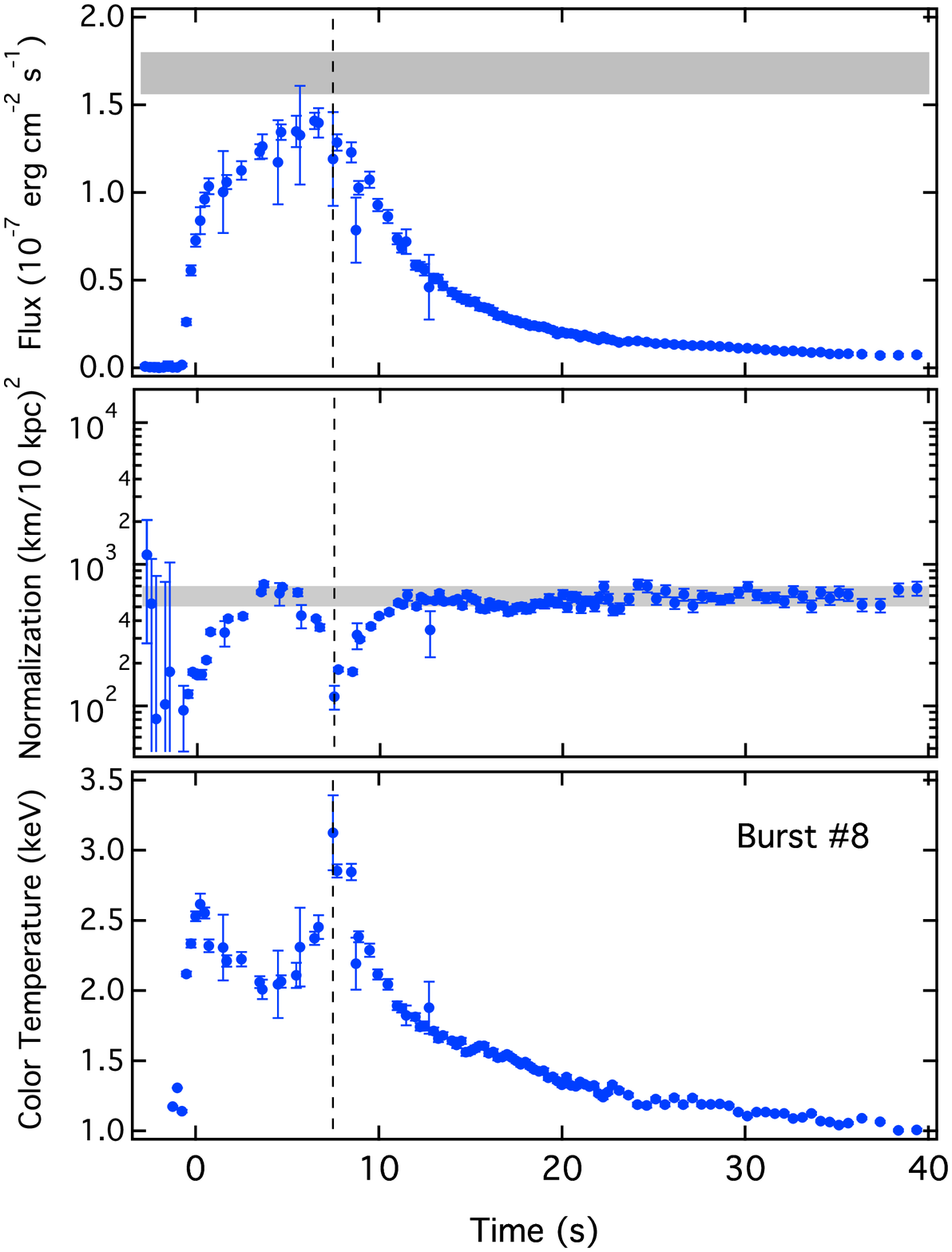}
\includegraphics[scale=0.35]{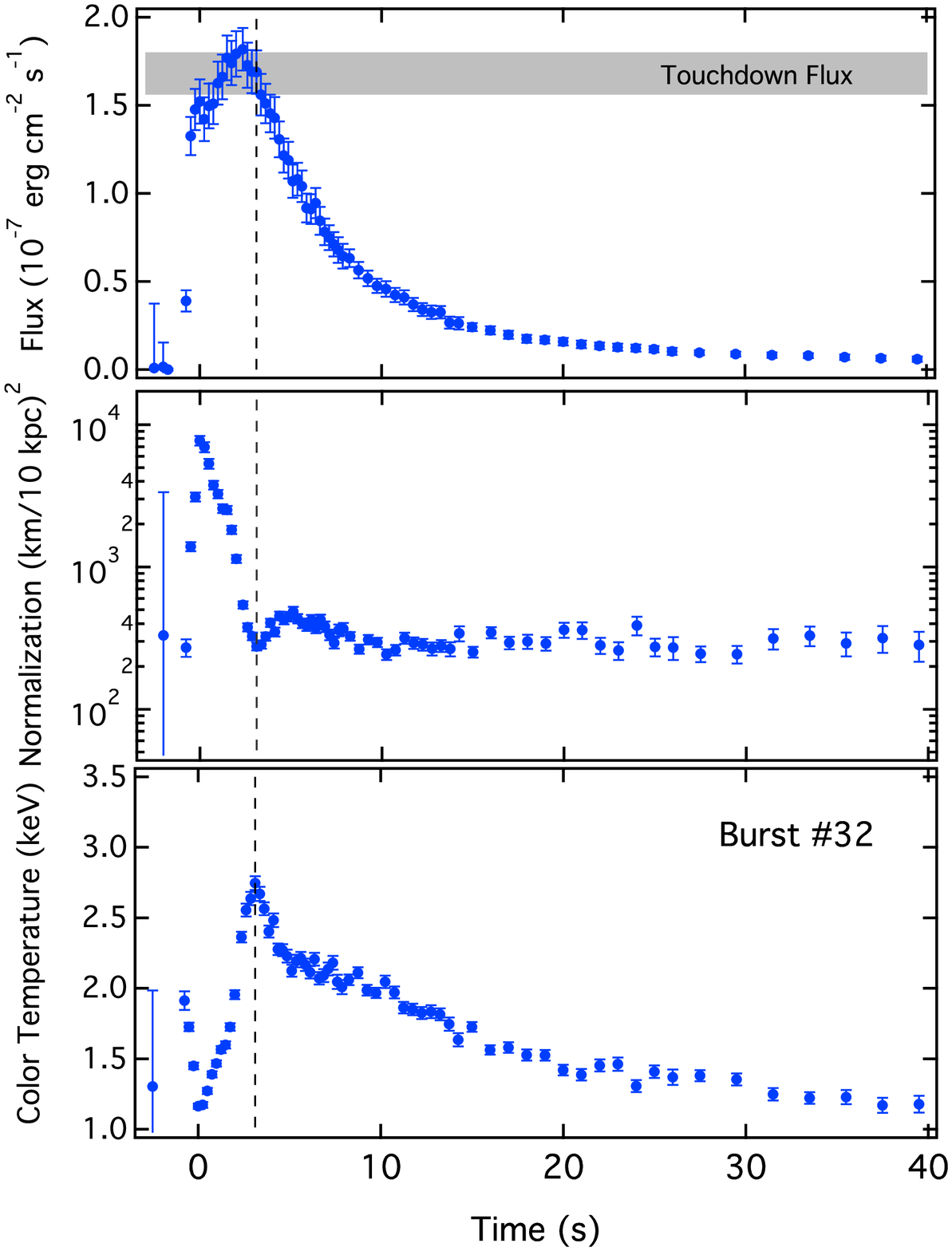}
\caption{The evolution of the spectral parameters of two bursts from
  \oeight.  The left panel shows one selected by \citet{poutanen2014}
  as a PRE burst, whereas the right panel shows a bona fide PRE burst
  according to \citet{guver2012a}. The blackbody normalization in the
  first case never exceeds the asymptotic value at the cooling tail
  and its touchdown flux is substantially below the touchdown flux of
  the second burst. This strongly argues against the classification of
  the bursts selected by \citet{poutanen2014} as PRE bursts. }
\label{fig:PRE_1608}
\end{figure*}

The identification of Eddington limited bursts relies exclusively on
finding evidence for photospheric radius expansion, where there is a
substantial increase in the inferred apparent radius of the
photosphere above the value measured in the cooling tail, coincident
with a decrease in the color temperature. Figure~\ref{fig:PRE_1608}
shows the evolution of the spectral parameters in two bursts from
\oeight. The burst on the left is representative of all the bursts
selected by \citet{poutanen2014} as PRE bursts that follow the
theoretical expectation. It is evident that even though the blackbody
normalization and the color temperature oscillate in the beginning of
the burst, the photosphere never expanded, i.e., the blackbody
normalization never exceeded the asymptotic value at the cooling tail.
In contrast, the burst on the right is a bona fide PRE burst and shows
the increase in the blackbody normalization during the radius
expansion episode by more than a factor of thirty. Furthermore, the
flux at the point that is identified as the touchdown on the left is
substantially ($\gtrsim 30\%$) smaller than the Eddington flux
measured at touchdown for the burst on the right. Even though the
method followed by \citet{poutanen2014} relies on measuring the
evolution of the color correction factor at flux levels within 10\% of
the Eddington limit, the selection procedure failed to identify bursts
that reach such flux levels.

\begin{figure*}
\centering \includegraphics[scale=0.35]{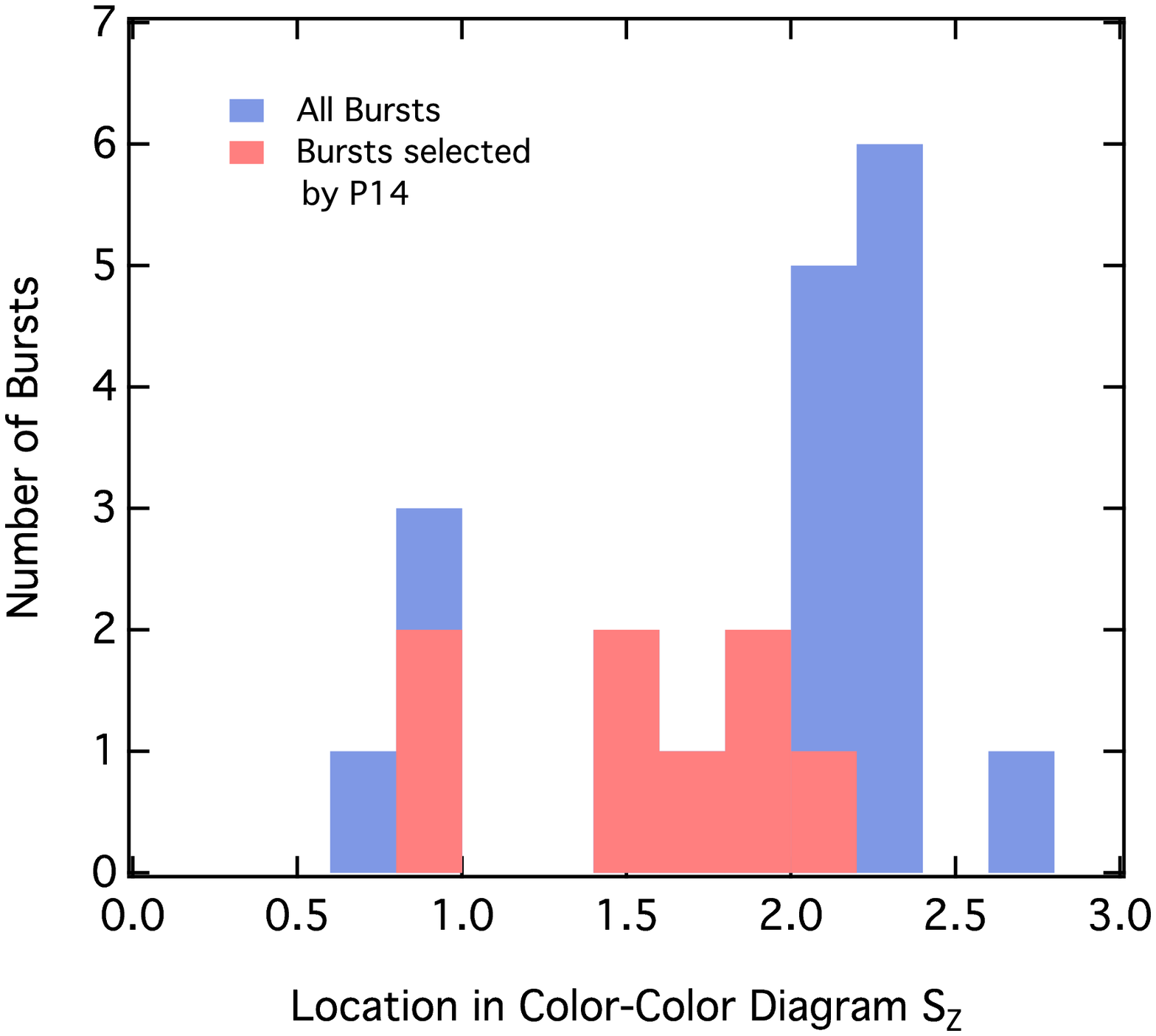}
\includegraphics[scale=0.35]{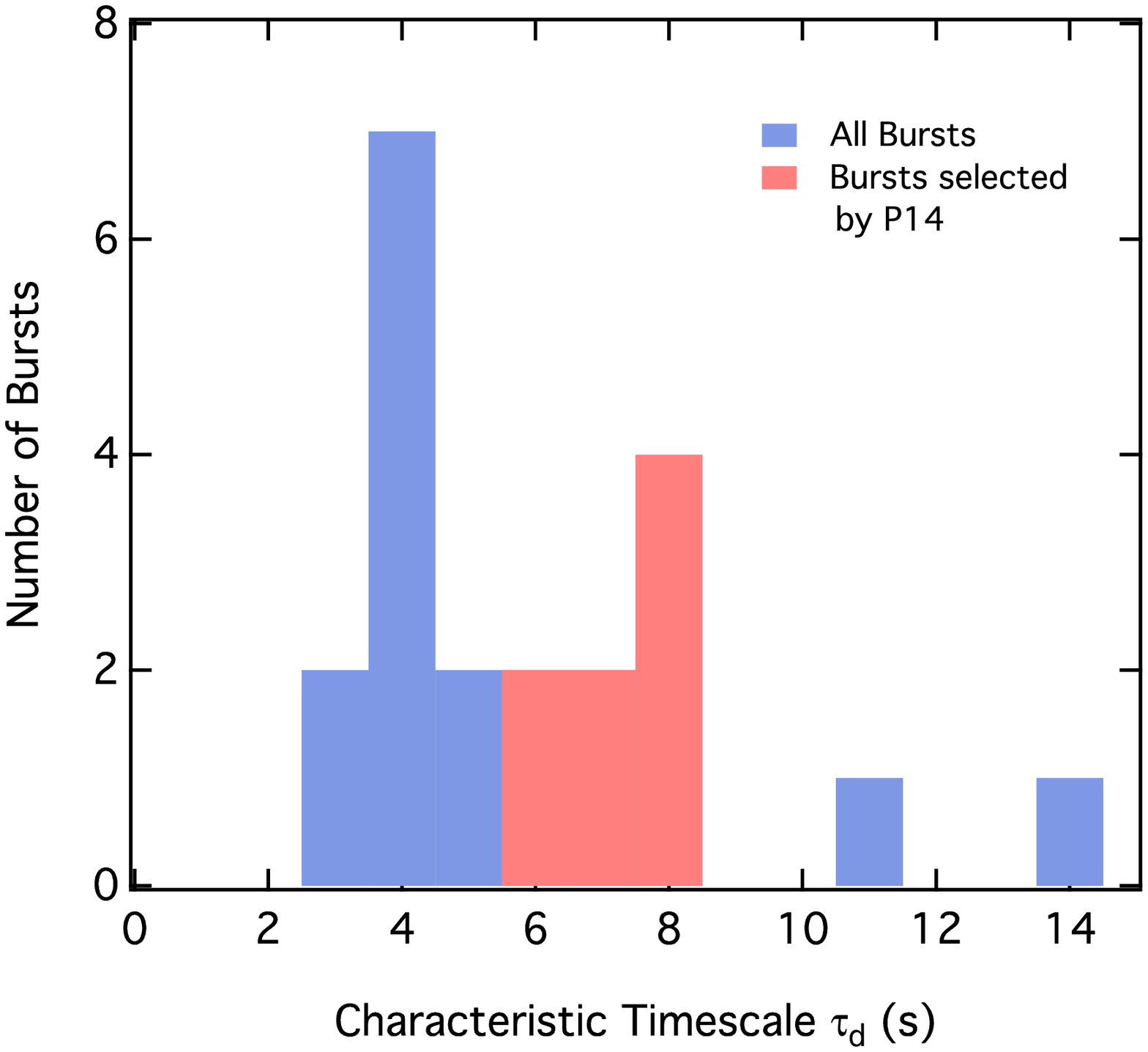}
   \caption{The distributions {\em (Left)\/} of hardness in the
     persistent emission (as measured by the location $S_{\rm z}$
     along the color-color diagram) and {\em (Right)\/} of the time
     between burst start and touchdown, for {\em (blue)\/} all the
     bursts of 4U~1608$-$52 analyzed by Poutanen et al.\ (2014) and
     {\em (red)\/} for those selected by them as being consistent with
     the theoretical predictions. The selected bursts are neither the
     longer bursts nor the ones that occur in the hardest state,
     contrary to earlier claims.}
\label{fig:burst_properties}
\end{figure*}

It is also worth emphasizing that, even though \citet{poutanen2014}
and \citet{kajava2014} characterize the selected bursts as long bursts
occurring in the hard state, implying an underlying physical reason
for this selection, their final sample represents neither the ones
that occur in the hard state nor the longest observed bursts. This can
be easily seen in Figure~\ref{fig:burst_properties} that compares the
distribution of {\it (i)} the burst locations in the color-color
diagrams and {\it (ii)} the time from burst start to touchdown between
the selected bursts and the entire sample for \oeight. Figures~3 and 
4 of \citet{kajava2014} corroborate these observations and challenge 
the interpretation that the theoretically motivated burst selection 
has an underlying physical mechanism (e.g., one that is related to 
the different states of the accretion flow). 

\section{Conclusions}

In this paper, we explored two approaches that have been proposed to
select thermonuclear burst data for the spectroscopic measurements of
neutron star radii. The first is the data-driven approach of
\citet{guver2012a,guver2012b} that used a Bayesian Gaussian mixture
algorithm to identify data outliers and measure the degree of
systematic uncertainty in the measurements. The second is the
theoretically motivated approach of \citet{poutanen2014} that required
selected bursts to follow trends expected from the bursting neutron
star atmosphere models of \citet{suleimanov2012}. The two approaches
led to non-overlapping data selections. Furthermore,
\citet{poutanen2014} and \citet{kajava2014} have used the
theoretically motivated model to argue that the data-driven selection
is not reliable because it is in conflict with the theoretical
expectations.

We showed that the theoretically expected trends between different
spectroscopic quantities, such as the blackbody normalization, the
color temperature, and the blackbody flux, get substantially smeared
in the data because of three effects. First, the inherent limitations
of the RXTE data do not allow resolving the rapid spectral evolution
that occurs at the end of the photospheric radius expansion episodes.
Second, even a mild scatter in the emitting area, e.g., due to uneven
burning, an evolving photosphere, partial obscuration of the surface,
and/or reflection off of the accretion disk, lead to a much larger
scatter in the measured spectroscopic quantities that mask the
theoretically expected trends. Finally, counting statistics lead to
correlated measurement uncertainties between the spectroscopic
quantities, further smearing any trends. We quantitatively showed that
these effects are at a level to prevent the detection of theoretical
trends in the majority of the burst data and preclude using these
criteria to argue against the data-driven selection methods.

Finally, we studied the sub-sample of bursts selected by
\citet{poutanen2014} for \oeight\ using the theoretically motivated
criteria without taking into account the data limitations discussed
above. We showed that, contrary to the implicit assumption in the
method, the bursts selected are not PRE bursts. Indeed, at no point
during any of these bursts does the radius of the photosphere exceed
the asymptotic radius measured during the cooling tails of the same
bursts. Furthermore, the inferred touchdown fluxes of these bursts is
30\% below the touchdown fluxes of the securely identified PRE bursts
that reach the Eddington limit. 

The quantitative analyses presented in this paper leads us to the
conclusion that this theoretically motivated selection procedure is
neither practical, given the limitations of the data, nor unbiased,
given that it selects bursts that are inconsistent with its own
assumptions. However, in the future, with the use of an X-ray detector
with a larger collecting area and significantly fewer limitations when
observing sources with high count rates, this could lead to useful
radius measurements. 

\acknowledgments

We thank all the participants of the ``The Neutron Star Radius''
conference in Montreal for helpful discussions and Gordon Baym,
Sebastien Guillot, and Craig Heinke for comments on the
manuscript. F\"O acknowledges support from NSF grant AST~1108753.  TG
acknowledges support from Istanbul University Project numbers 49429
and 48285.

\bibliography{/Users/fozel/Dropbox/Publications/ARAA/ARTICLE/feryal}

\end{document}